\documentclass[aps,physrev,rreprint,12pt,groupedaddress]{revtex4-2}
\usepackage[margin=0.9in]{geometry}
\usepackage{setspace}
\usepackage{titlesec}

\usepackage{graphicx}
\usepackage{dcolumn}
\usepackage{bm}
\usepackage[normalem]{ulem}
\usepackage{amsmath,amssymb,amsfonts,xcolor}%

\begin{document}

\title{Confined drying of a binary liquid mixture droplet: \\ A quantitative interferometric study under humidity control}

\author{Ole Milark$^{1}$}
\author{Jean-Baptiste Salmon$^{2}$}

\author{Benjamin Sobac$^{1}$}
\email[]{benjamin.sobac@cnrs.fr}

\affiliation{$^{1}$ CNRS, Université de Pau et des Pays de l'Adour, LFCR, 64600 Anglet, France}

\affiliation{$^{2}$ CNRS, Syensqo, Université de Bordeaux, LOF, 33600 Pessac, France}

\begin{abstract}
We present a methodology that combines Mach-Zehnder interferometry, a custom relative humidity (RH) controlled chamber, and a confined two-dimensional droplet geometry to enable precise investigations of drying of complex fluids and the associated transport mechanisms. This approach is applied to a model binary mixture, water-glycerol, the concentration-dependent thermodynamic and transport properties of which are relatively well documented. High-resolution interferometric imaging (6~\textmu m\,pixel$^{-1}$,  1~frame\,s$^{-1}$) allows simultaneous measurement of drying kinetics and internal concentration fields with $\pm 0.5\%$ accuracy, characterized here over a wide range of RH (25--95\%), and thus P\'eclet numbers. The experimental results closely match  a quasisteady, isothermal model of vapor-diffusion-controlled evaporation coupled to  diffusion within the droplet. These data enable  extraction of both the concentration-dependent mutual diffusion coefficient $D(\varphi)$ and the water chemical activity $a_w(\varphi)$ over almost the entire range of glycerol volume fraction $\varphi$, even from a single low-RH experiment. While $a_w(\varphi)$ agrees well with literature values, our measurements yield a consistent fit for $D(\varphi)$. Complementary experiments with fluorescence microscopy confirm that buoyancy-driven convection, although present, remains negligible, so that mass diffusion dominates solute transport in this confined geometry. The overall agreement validates the methodology, demonstrating its robustness as a quantitative framework for probing drying dynamics and transport in complex fluids, with broad applicability to controlled evaporation studies.

\end{abstract}


\maketitle

\section{INTRODUCTION}

The drying of solutions or dispersions is a key step in a wide range of processes, from inkjet printing~\cite{derby_inkjet_2010,teichler_inkjet_2013,lohse2022fundamental}, liquid-based coating techniques~\cite{grosso_how_2011,butt_thin-film_2022}, and spray drying \cite{lee_formation_2010,Lintingre,boel2020unraveling} (widely used in food and pharmaceuticals~\cite{vehring2008pharmaceutical,ZIAEE2019300,schuck2016recent, dantasInnovationsSprayDrying2024}) to conservation of painted artworks~\cite{chelazziNanoparticlesGelsBreakthrough2023,pauchard2020craquelures}, fabrication of solar cells and battery electrodes~\cite{krebsFabricationProcessingPolymer2009,zhangReviewLithiumIonBattery2022,Li2022}, and numerous biomedical or diagnostic applications~\cite{sefiane2021patterns,pal_drying_2023}. In all these processes, the drying kinetics and the resulting final morphology (deposits, films, or particles) result from subtle interactions between thermodynamics driving solvent evaporation and nonequilibrium phenomena (rheology and mass transport), which affect the evolving concentration fields of solutes/particles in solution~\cite{routhDryingThinColloidal2013,bacchin2018drying}. These questions are even relevant in the context of infectious-disease transmission, for instance, to understand the drying dynamics of suspended respiratory droplets~\cite{Merhi2022,Haut_2025} or the spatial distribution of viral particles within the resulting dried residues~\cite{Martinez}.

To study these couplings, model geometries that enable precise control of evaporation and transport phenomena are required. These include drying in microfluidic channels~\cite{Daubersies:13}, capillaries or Hele–Shaw cells with transverse dimensions below $100~\text{\textmu m}$~\cite{Dufresne:06,Hooiveld2023,Goehring2017,Roger2022,Raju2024}, as well as drying of droplets in various configurations:  confined two-dimensional (2D) droplets~\cite{clement_evaporation_2004,Boulogne:13}, or three-dimensional droplets --- sessile~\cite{bodiguel2010imaging,sefiane2014patterns, brutin2015droplet, li2019gravitational,thayyil2022evaporation,gelderblom2022evaporation}, levitated or suspended~\cite{Lintingre,tsapis2005onset,davies2012time,fu2012single,krieger2012exploring,yarin2002evaporation,archer2020drying}.

Analytical techniques that are spatially and temporally resolved are required to measure the local concentration of solutes/particles during drying and/or the flows involved. These techniques include  observations with conventional microscopy  [bright field (BF), fluorescence, confocal] but also advanced methods to measure concentration fields, such as Raman microspectroscopy~\cite{Daubersies:13,Roger2022}, thermospectroscopic infrared imaging~\cite{lehtihet_thermospectroscopic_2021}, magnetic resonance imaging~\cite{erb2025visualization}, small-angle x-ray scattering~\cite{Goehring2017}, or optical coherence tomography~\cite{abe_dynamics_2024}.

In previous work, we investigated the drying of different complex fluids in  a confined 2D droplet, including aqueous copolymer solutions~\cite{daubersies_confined_2012}, and charged colloidal dispersions~\cite{loussert_drying_2016,sobac_collective_2020}. The geometrical confinement of such experiments leads to isothermal evaporation conditions, with mass transport dominated by diffusion (no capillary/Marangoni flows, limited free convection). Measurements of concentration fields were initially performed using Raman microspectroscopy, however its low accuracy and temporal resolution prevented the extraction of precise data for the complex fluids under study~\cite{daubersies_confined_2012,loussert_drying_2016}. More recently~\cite{sobac_collective_2020}, we demonstrated  that Mach-Zehnder interferometry provides significantly more accurate measurements of concentration during the  drying of confined 2D droplets, with high spatial and temporal resolution. This enabled us, in particular, to extract the mutual diffusion coefficient of charge-stabilized dispersion over a wide concentration range, evidencing the role played by colloidal interactions on this transport coefficient. Nevertheless, none of these measurements were performed on a model system to assess the validity of the approach or test its precision. Also,  relative humidity (RH) was not controlled in such experiments, while RH not only impacts the drying kinetics of aqueous complex fluids, but also the concentration level reached at the end of drying when the chemical activity of water $a_w$ depends on concentration, as is the case for concentrated solutions of polymers or small molecules.

In the present work, we built a Mach-Zehnder interferometer coupled with a custom-made RH-controlled chamber to thoroughly investigate 
the drying of a confined 2D droplet composed of
a model binary mixture: water-glycerol, a binary liquid system commonly employed in studies on drying and evaporation~\cite{Raju2024,Li2022,thayyil2022evaporation,davies2012time}. Thermodynamic properties (water chemical activity $a_w$, density $\rho$) and most of the transport coefficients (viscosity $\eta$, mutual diffusion coefficient $D$) of this mixture  are well documented, although some discrepancies exist in data of the literature regarding $D$~\cite{bouchaudy_steady_2018}. We then demonstrate that the high accuracy of the Mach-Zehnder interferometer ($\pm 0.5\%$) and its resolution (6~\textmu m\,pixel$^{-1}$,  field of view 8.7$\times$6.5~mm$^2$, 1~frame\,s$^{-1}$), combined with precise RH control during long-duration drying experiments 
($\pm 3.5\%$), enable a detailed and precise characterization of the drying of this binary mixture. At the same time, these experiments allow us to provide estimates of the mutual diffusion coefficient $D$ over glycerol volume fractions ranging from $\varphi \simeq 0.2$ to 0.9.

The manuscript is organized as follows. Section~\ref{sec:Model} summarizes the main features of the drying of a confined 2D droplet in the case of a binary mixture, water $+$ nonvolatile solute. Then, Sec.~\ref{sec:MM} details the physical properties of the water-glycerol system, describes the experimental setups including the interferometric and microscopic imaging techniques, and outlines the image processing methods employed. In Sec.~\ref{sec:RD}, we present the experimental results, along with a comprehensive comparison with a transport model from which both water chemical activity $a_w$  and mutual diffusion coefficient $D$ are extracted from experiments and compared to literature data. Finally, we conclude by highlighting the opportunities offered by the developed techniques for the study of other complex fluids.

\section{\label{sec:Model} CONFINED DRYING OF A 2D DROPLET}

\begin{figure}
    \centering
    \includegraphics{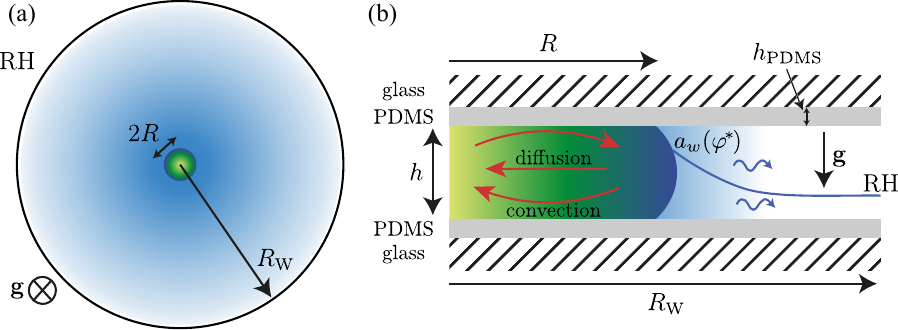}
    \caption{\label{fig:model} 2D confined drying droplet configuration. (a) A liquid droplet of radius $R(t)$ dries while confined between two circular polydimethylsiloxane (PDMS)-coated glass wafers of radius $R_\text{W}$. The wafers are separated by a fixed gap of height $h$, maintained by spacers, forming a quasi-2D geometry with $h \ll R_\text{W}$. (b) Schematic cross-sectional view of the confined droplet and the main physical processes involved during its drying. The two color maps represent the two concentration gradients, in the droplet and in the gas phase.}
\end{figure}

In this work, we used the confined drying configuration illustrated schematically in Fig.~\ref{fig:model}(a), originally introduced by Clément and Leng~\cite{clement_evaporation_2004}. A small droplet of an aqueous sample, of volume $\simeq$1$-$2~\textmu L, is squeezed between two circular wafers of radius $R_\textrm{W}$, separated  by a distance $h$ using  spacers. In most experiments~\cite{daubersies_confined_2012,loussert_drying_2016,sobac_collective_2020}, wafers have a diameter of 3~in. ($R_\textrm{W} = 38.1~$mm), $h$ ranges from $100$ to $200~$\textmu m, and the initial droplet radius ranges from $R_0 = 1$ to 2~mm. The evaporation of water in this confined cell ultimately leads to an axisymmetric shrinkage of the droplet, with a typical drying time of a few hours under ambient conditions. 

The drying kinetics for a binary mixture, solvent $+$ nonvolatile solute, has been theoretically described in Ref.~\cite{daubersies_evaporation_2011}. Below we briefly summarize the main elements of such an axisymmetric model in the case of an aqueous solution (see also Ref.~\cite{sobac_collective_2020}). Drying occurs in  isothermal conditions through the quasistatic diffusion of vapor from the droplet to the edge of the cell at $r=R_\textrm{W}$ (see Fig.~\ref{fig:model}).
The molar   concentration of vapor $C$  thus obeys the Laplace equation $\Delta C = 0$ for $R(t) < r < R_\text{W}$, $R(t)$ being the droplet radius at time $t$. After integration, $C$ is given by
\begin{equation}
   C(r,t) = C_\text{sat} \left(a_w^* 
   -(a_w^*- \text{RH}) \frac{\ln[r/R(t)]}{\ln[R_\text{W}/R(t)]}\right),
\label{eq:aw_r}
\end{equation}
in which $C_{\textrm{sat}}$ is the water concentration in the gas phase at saturation, $\text{RH}$ is the relative humidity outside the cell, and $a_w^*$ is the water chemical activity at the liquid-air interface.
Conservation of volume ultimately leads to the drying kinetics  of the droplet:
\begin{equation}
    \frac{\text{d}R(t)}{\text{d}t} = \frac{D_w V_w C_{\textrm{sat}}}{R(t) \ln{[R(t)/R_{\textrm{W}}]}}(a_w^* - \textrm{RH}),
\label{equ:evolution_alpha}
\end{equation}
where $D_w$ is the diffusion coefficient of water vapor in air and $V_w$ is the molar volume of liquid water. In case of a pure water droplet, or for a dilute mixture, $a_w^* = 1$ and  Eq.~\eqref{equ:evolution_alpha} admits the following analytical solution:
\begin{equation}
    \frac{t}{\tau_f} = 1 - \alpha \frac{ \ln(\beta\alpha) -1}{\ln(\beta)-1},
\label{equ:analytical}
\end{equation}
with the normalized droplet area $\alpha = [R(t)/R_0]^2$, $\beta = (R_0/R_\text{W})^2$, $R_0 = R(t=0)$, and $\tau_f$ given by 
\begin{equation}
    \tau_{f} = \frac{R_{\textrm{0}}^2}{4 D_w V_w C_{\textrm{sat}} (1-\textrm{RH})} [1 - \textrm{ln}(\beta)].
\label{eq:tau_f}
\end{equation}
$\tau_f$  corresponds to the drying time of a pure water drop, i.e., $\alpha(t\to \tau_f) \to 0$.
At $T=22^\circ$C, $C_\textrm{sat} \simeq 1.08$~mol\,m$^{-3}$, $V_w \simeq 1.8 \times 10^{-5}$~m$^3$\,mol$^{-1}$, and  $D_w \simeq 2.5 \times 10^{-5}$~m$^2$\,s$^{-1}$, so that $\tau_f \simeq 2$~h for a droplet of radius $R_0 = 1$~mm drying at $\mathrm{RH}=40\%$.
This simple estimate also fully justifies the quasistatic approximation, since  $\tau_f \gg R_w^2/D_w \simeq 1$~min (see Ref.~\cite{daubersies_evaporation_2011} for details).

In case of an ideal  binary mixture, i.e., no change of volume with the concentration, and when internal flows due, for instance, to free convection are negligible, the solute volume fraction $\varphi$  is governed by diffusion only and given by the axisymmetric diffusion equation
\begin{equation}
    \frac{\partial \varphi}{\partial t} = \frac{1}{r} \frac{\partial}{\partial r} \left( r D(\varphi) \frac{\partial \varphi}{\partial r} \right),
\label{eq:diff}
\end{equation}
where $D(\varphi)$ is the mutual diffusion coefficient of the mixture.
The nonvolatility of the solute results in the  boundary condition
\begin{equation}
    -D(\varphi)\frac{\partial \varphi}{\partial r}\bigg|_{r=R(t),t} = \varphi[R(t),t]\frac{\text{d}R}{\text{d}t}.
\label{eq:nonvo_bc}
\end{equation}
The average solute volume fraction in the droplet $\langle\varphi(t)\rangle$  is directly related to the normalized area $\alpha$ through the global solute mass balance
\begin{equation}
\label{eq:avg_phi_ot}
    \langle\varphi(t)\rangle=\frac{1}{\pi R(t)^2}\int_0^{R(t)} 2 \pi r\varphi(r,t) \mathrm{d}r=\frac{\varphi_0}{\alpha(t)}\, , \end{equation}
where $\varphi_0 = \varphi(r,t = 0)$ is the initial solute volume fraction in the droplet, assumed uniformed for simplicity. 
Equations~\eqref{equ:evolution_alpha}, \eqref{eq:diff} and \eqref{eq:nonvo_bc} can be solved to predict the evolution of the solute concentration inside the droplet during drying assuming local thermodynamic equilibrium at the liquid-air interface, i.e., $a_w^* = a_w(\varphi^*)$ where $\varphi^* = \varphi[r=R(t),t]$ and $a_w(\varphi)$ is the water chemical activity of the mixture. We will return later to the strong assumption of negligible internal flows, particularly with regard to the unavoidable free convection  driven by the drying-induced density gradients.

The above model naturally leads to the definition of the P\'eclet number:
\begin{equation}
\mathrm{Pe} = \frac{R_{0}^{2}}{D_0 \tau_f},
\label{eq:Pe}
\end{equation}
in which $D_0 = D(\varphi \to 0)$.  $\mathrm{Pe}$ compares the typical timescale of diffusion across the droplet $\sim$$R_0^2/D_0$ with the  drying time $\tau_f$. For $\mathrm{Pe} \ll \mathcal{O}(1)$, diffusion flattens the concentration gradients and the droplet dries almost homogeneously. When $\mathrm{Pe} \gg \mathcal{O}(1)$, strong concentration gradients form at the receding liquid-air interface.
Because of Eq.~\eqref{eq:tau_f}, $\mathrm{Pe}$ reduces to
\begin{equation}
    \mathrm{Pe} = \frac{4 D_w V_w C_{\textrm{sat}} (1-\textrm{RH})}{D_0 [1 - \ln(\beta)]} = \frac{\tilde{D}}{D_0},
    \label{eq:Pedevelop}
\end{equation}
and depends only weakly  on  $R_0$ through the term $\ln(\beta)$. For initial droplet radius in the range $R_0 = 1$--$2$~mm, $\ln(\beta) \simeq -[5.8$--$6.5]$ and $\tilde{D}$ varies from 0.4 to $2\times 10^{-10}$~m$^2$\,s$^{-1}$ for $\text{RH}$ varying from 85 to 30\%.

Appendix~\ref{sec:App_Model} provides the dimensionless form of the model given by Eqs.~\eqref{equ:evolution_alpha}, \eqref{eq:diff}, and \eqref{eq:nonvo_bc}, along with the numerical methods used to solve it. These dimensionless equations  reveal the key parameters governing the confined drying experiment: $\beta$, $\tau_f$, Pe, and $\varphi_0$.

\section{\label{sec:MM}MATERIALS AND METHODS}

\subsection{Water-glycerol mixture}

The binary liquid mixture considered consists of distilled water (Milli-Q, 18.2 M$\Omega$ cm) and glycerol (Sigma-Aldrich G5516, molecular biology grade, purity $\geq$99.0\%). The solutions are prepared by mass dilution using a precision scale to achieve glycerol mass fractions in the range $w=0.04$ to 0.09 with an accuracy of $\simeq$0.1\%.  The water-glycerol mixture can be considered almost ideal, meaning that the total volume remains virtually unchanged upon mixing~\cite{bouchaudy_steady_2018}. In that case, one can define the  glycerol volume fraction  from $w$ as 
\begin{equation}
\label{eq:vaphivsw}
\varphi = \frac{\rho}{\rho_{g}}\, w,
\end{equation}
where $\rho$ is the density of the mixture, following  the linear variation
\begin{equation}
\label{equ:density}
\rho(\varphi) \simeq \rho_{w}(1+\beta\varphi),
\end{equation}
with $\rho_{w}=997.8$~kg\,m$^{-3}$ the density of pure water at $T=22^{\circ}$C, $\beta =  0.2623$ the solutal expansion coefficient of the mixture, and $\rho_g = \rho_w(1+\beta)$. This yields  water-glycerol mixtures with initial glycerol volume fractions ranging from $\varphi_0 = 0.0320$ to $0.0727$ with an accuracy of $ \pm 0.0003$ due to errors in mass measurements.

\begin{figure}
\includegraphics{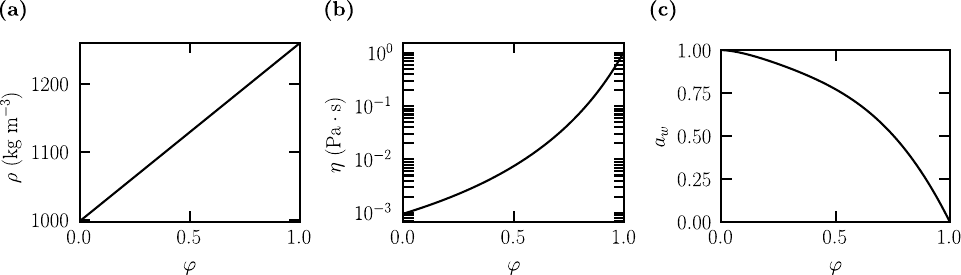}
\caption{\label{fig:prop_gly_water} Physical properties of the water-glycerol mixture at $T=22^{\circ}$C. (a) Density $\rho$, (b) viscosity $\eta$, and (c) water chemical activity $a_w$ as a function of the glycerol volume fraction $\varphi$. Panel (a) corresponds to Eq.~\eqref{equ:density}, panel (b) is the empirical correlation of Ref.~\cite{cheng_formula_2008}, and panel (c) corresponds to Eq.~\eqref{equ:bouchaudy_activity}.}
\end{figure}

Several physical properties of the water-glycerol mixture relevant for our study are shown in Fig.~\ref{fig:prop_gly_water} as a function of $\varphi$ at $T=22^{\circ}$C. The linear variation of density predicted by Eq.~\eqref{equ:density} is shown in Fig.~\ref{fig:prop_gly_water}(a). Figure~\ref{fig:prop_gly_water}(b) displays the empirical correlation derived by \citet{cheng_formula_2008} that accounts for the nonlinear dependence of viscosity $\eta$ on $\varphi$ over the entire concentration range. Finally, Fig.~\ref{fig:prop_gly_water}(c) displays the empirical relation of \citet{bouchaudy_steady_2018}:
\begin{equation}
a_{w}(\varphi) = (1-\varphi)\left(0.819\,\varphi^3 - 0.0971\,\varphi^2 + 0.831\,\varphi + 1\right),
\label{equ:bouchaudy_activity}
\end{equation}
fitting experimental data for the water chemical activity of aqueous glycerol solutions~\cite{kirgintsev_water_1962,zaoui-djelloul-daouadji_vapor-liquid_2014,ninni_water_2000,marcolli_water_2005}.
To date, there is only a limited amount of experimental data on the mutual diffusion coefficient $D(\varphi)$ of the water-glycerol mixture~\cite{nishijima_diffusion_1960,ternstrom_mutual_1996,derrico_diffusion_2004,rashidnia_measurement_2004,bouchaudy_steady_2018}. These data will be used later for comparison with our results.

\subsection{Confined drying experiments under a humidity-controlled environment}

The confined drying experiment consists of depositing, using a micropipette, a droplet of the water-glycerol mixture with an initial volume of $1–2~\text{\textmu L}$ between two glass wafers (3 in. in diameter, $R_\text{W}=38.1~$mm, $500 \pm 20~\text{\textmu m}$ thick) and allowing it to dry naturally in air.  The two wafers are separated by a constant height $h = 150 \pm 1~\text{\textmu m}$ using three  pieces of glass cover slides as spacers. Each spacer is small (a few mm$^2$) and positioned far from the droplet, equidistantly along the wafer edge, so as not to affect evaporation. Under these conditions, the initial droplet radius ranges from $R_0 = 1.45$ to 2.06~mm.
Thin layers of polydimethylsiloxane (PDMS, Sylgard 184, thickness $h_\text{PDMS} \simeq 24~\text{\textmu m}$) are deposited on the
two glass wafers by spin coating followed by thermal  cross-linking. These hydrophobic coatings prevent the receding meniscus from sticking during the drying of the droplet, with an expected contact angle of $100^{\circ}$--$110^{\circ}$ \cite{zhangContactAngleMeasurement2021}. In previous studies employing this configuration~\cite{daubersies_confined_2012,loussert_drying_2016,sobac_collective_2020}, the PDMS layers have always been considered to have no effect on droplet drying, despite PDMS being known to absorb moisture when exposed to vapor or liquid water~\cite{Harley2012}. We justified in the present work with combined experiments and modeling (see Appendix~\ref{ss:influence_of_PDMS}) that this is indeed the case for a thin thickness $h_\text{PDMS} \simeq 24~\text{\textmu m}$, but no longer at larger thicknesses where slight deviations from the model described by Eq.~\eqref{equ:evolution_alpha} occur.

The confined drying experiments are performed inside a custom-built humidity-controlled chamber, designed following the setup of Ref.~\cite{boulogne_cheap_2019}, allowing precise control of the air relative humidity between $\textrm{RH}=25$ and $95\%$, with a sensor precision of $\pm3.5\%$. Experiments are conducted at room temperature ($T=22^{\circ}$C), as monitored by an internal thermocouple, and under ambient atmospheric pressure. Two chamber volumes are designed to accommodate the geometrical constraints of the two experimental setups used in this study, as described in the following sections: one of $1270~\text{cm}^3$ for interferometry and another of $130~\text{cm}^3$ for optical microscopy, with respective air inlet flow rates of 40 and 2~L\,h$^{-1}$.

In the experiments, RH in the chamber is set to the desired value prior to each run, with a typical equilibration time of $1$--$20$~min. Particular care is then paid to minimize the delay between droplet deposition on the lower glass wafer and the start of the experiment ($t = 0$). This includes (i) confining the droplet with the upper wafer, (ii) transferring the confined cell to the chamber using a manual drawer [see Fig.~\ref{fig:experimental_setup}(a)], and (iii) centering the droplet in the camera’s field of view using a manual stage. This preparation sequence never exceeds 30~s, ensuring that solvent evaporation, and thus any variation in the initial concentration, remains negligible. We also verified that this brief preparation period, during which the drawer is opened momentarily, has no effect on the relative humidity imposed in the chamber.

\subsection{Mach-Zehnder interferometry}
\label{sec_MZI}

\subsubsection{Experimental setup}

\begin{figure}
\includegraphics{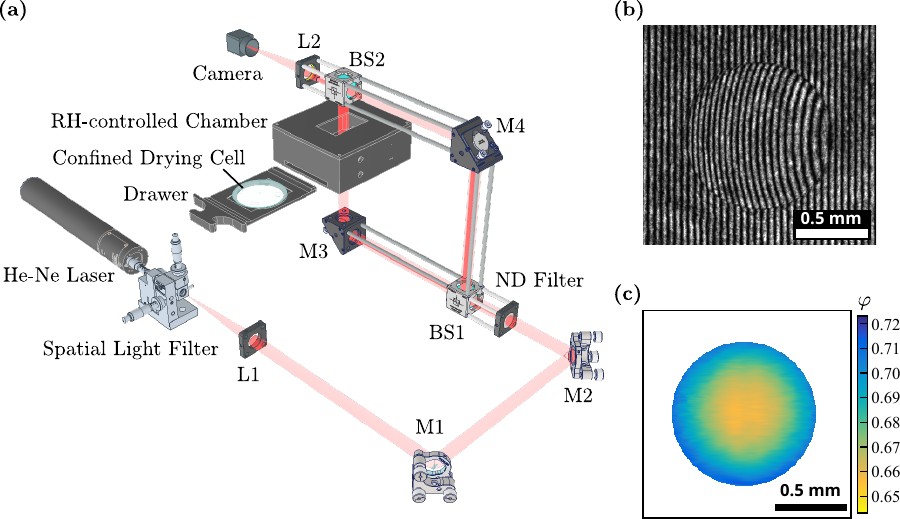}
\caption{\label{fig:experimental_setup} Experimental setup for combined measurements of drying kinetics and concentration fields during drying. (a) Mach-Zehnder interferometer coupled with the 2D confined droplet cell inside a humidity-controlled chamber. (b) Typical raw interferogram of a confined drying droplet of a water-glycerol mixture recorded during an experiment. (c) Corresponding concentration field inside the droplet obtained after postprocessing of the interference fringes.}
\end{figure}

\begin{figure}[ht!]
\includegraphics{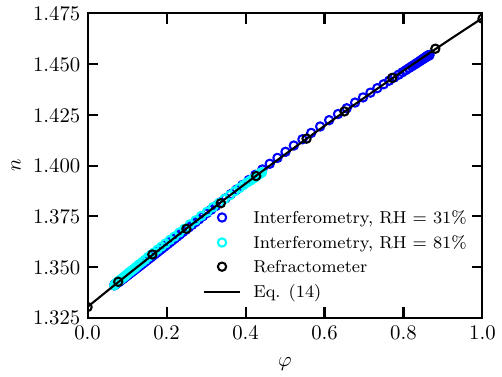}
\caption{\label{fig:refr_index_gly_water}Refractive index of the water-glycerol mixture as a function of the glycerol volume fraction at $\lambda=632.8$~nm and $T=22^{\circ}$C. Light and dark blue symbols represent autocalibration curves from interferometry experiments at two RH values, while black symbols correspond to independent measurements performed using a refractometer. Data are well described by the second-degree polynomial fit provided in Eq.~\eqref{eq:refr_index}.}
\end{figure}

The humidity-controlled chamber containing the confined drying droplet is housed inside a Mach-Zehnder interferometer, as shown in Fig.~\ref{fig:experimental_setup}(a). This optical setup enables the visualization and measurement of the 2D concentration field within the drying droplet, making use of the concentration dependence of its refractive index $n$ (see Fig.~\ref{fig:refr_index_gly_water}). A stabilized He-Ne laser (mks Newport N-STP-912, wavelength $\lambda=632.8$~nm, output power 1 mW) provides the coherent light source. The beam is expanded and cleaned using a spatial filter (Newport M900) equipped with a $20\times$ microscope objective and a 20~\textmu m pinhole, then collimated to a diameter of 18~mm by lens L1 ($f = 150$~mm). Its intensity is adjusted with a neutral density filter before being split by the first beam splitter (BS1) into a reference beam and a measurement beam, the latter passing through the confined drying droplet. The two beams are recombined at the second beam splitter (BS2) and imaged onto a CMOS camera (IDS U3-3040CP-M-GL Rev.2.2, $1456\times 1088$ pixels).  Lens L2 ($f = 50 \textrm{ mm}$) is used to focus the camera onto the confined droplet. Due to the refractive index differences in the droplet, the optical paths of the two beams are different

\noindent and an interference pattern is recorded. A typical interferogram is shown in Fig.~\ref{fig:experimental_setup}(b). Note that the small circular black strip visible in the interferogram corresponds to the liquid-air interface, the meniscus shape deflecting the measured beam out of reach of the objective. Moreover, for image processing purposes, the interferometer is set to generate a homogeneous system of vertical fringes before the droplet is deposited, as still observed outside of the droplet in the interferogram.
The typical acquisition rate in our experiments is one frame every 10~s, with a spatial resolution of approximately $6~$\textmu m pixel$^{-1}$. 

\subsubsection{Image processing}

The finite-fringe interferograms are first analyzed to  extract the phase field $\phi$, following the methodology fully detailed in Ref.~\cite{sobac_collective_2020}. This procedure includes classic Fourier-transform analysis using standard literature algorithms \cite{takeda_1982,kreis_1986}, spatial phase unwrapping via the algorithm of Herr\'{a}ez \textit{et al.}~\cite{herraez_fast_2002}, and custom temporal unwrapping and phase correction as described in Ref.~\cite{sobac_collective_2020}. 
A minor adjustment is introduced for droplet meniscus detection, as operating in Fourier space allows, in parallel with fringe analysis, the generation of a fringe‑cleaned image from which the droplet contour can be automatically detected. 
This step facilitates both the characterization of drying kinetics via contour analysis and droplet masking required for phase processing. Details of these image-processing steps are provided in Supplemental Material~\cite{Supp_Mat}. 
From the contour, the droplet area \(A(t)\) is extracted using a standard edge-detection algorithm (Canny edge detection from Python package \texttt{OpenCV}). The equivalent droplet radius and the average glycerol volume fraction are then computed, respectively, as $R(t) = \sqrt{A(t)/\pi}$, assuming circular symmetry, and $\langle \varphi(t) \rangle=\varphi_0\, [R_0/R(t)]^{2}$ from the solute mass conservation Eq.~\eqref{eq:avg_phi_ot}.

Once the phase field $\phi$ is obtained, it is converted to the refractive index using the relation
\begin{equation}
    \Delta n = \frac{\lambda \Delta \phi}{2 \pi h},
\end{equation}
valid for Mach-Zehnder interferometry in a Hele-Shaw cell~\cite{parimalanathan_2026}. Since this relation is relative, an initial refractive-index value must first be assigned: the refractive index within the droplet in the first image is taken to be homogeneous and equal to that of the initial mixture, an assumption justified by the analysis of the profiles at $t=0$ (see Fig.~\ref{fig:profiles_comparison}). 

The full transformation of refractive index $n$ into glycerol concentration $\varphi$ then follows from a calibration curve of $n(\varphi)$, obtained independently using a refractometer (Abbemat, Anton Paar) at $T=22^{\circ}$C for $\lambda=632.8~$nm. The calibration curve, displayed in Fig.~\ref{fig:refr_index_gly_water}, includes these measurements, as well as autocalibration curves extracted from two experiments performed at $\mathrm{RH}=31$ and 81\%, respectively, with a full analysis presented later (see Fig.~\ref{fig:profiles_comparison}). Indeed, from the image processing, one can extract the average refractive index $\langle n \rangle$ as a function of the average glycerol volume fraction $\langle\varphi\rangle$ in the droplet during  drying. All these data collapse and are nicely fitted by the second-degree polynomial law
\begin{equation}
n(\varphi) = 1.33034 + 0.15825\; \varphi - 0.01587\;\varphi^2.
\label{eq:refr_index}
\end{equation}
This calibration step finally leads to the spatial distribution of glycerol volume fraction shown in Fig.~\ref{fig:experimental_setup}(c). Given its axisymmetric nature, the 2D volume fraction fields are azimuthally averaged to obtain the radial profiles $\varphi(r,t)$ presented in Sec.~\ref{sec:RD}. 
Variations of $n$ with $\varphi$ for the system under study are close to the one investigated in \citet{sobac_collective_2020}, so that we can deduce that 
Mach-Zehnder interferometry provides volume fraction fields and radial profiles with an absolute accuracy better than 0.5\% in the glycerol volume fraction.

\subsection{Bright-field and fluorescence microscopy}

Classical BF and fluorescence microscopy are also employed with the humidity-controlled chamber to complement the characterization of confined drying, providing detailed information on drying kinetics and internal flows, respectively. A multimode inverted microscope (Nikon Ti2-U) equipped with a 4$\times$ objective (Nikon CFI Plan Fluor, NA = 0.13) and a sCMOS camera (Hamamatsu ORCA Flash 4 LT 3, 2048$\times$2028 pixels) is used for both imaging modes, focusing at a plane located at $z=h/2$. Recorded images have a spatial resolution of $1.625~$\textmu m per pixel.

For BF imaging, the drying droplet is homogeneously illuminated in transmission mode by a white LED (CoolLED Pt-100-WHT). Although the droplet drying kinetics can be extracted from interferometric images (see Sec.~\ref{sec_MZI} or Ref.~\cite{sobac_collective_2020}), BF imaging is used for edge detection of the droplet contour when interferometric measurements are not required. BF images are then processed using the same automatic contour detection method as described above to extract $A(t)$, $R(t)$, and $\langle\varphi(t)\rangle$.

For fluorescence imaging, the confined droplet is illuminated in reflection mode by a green LED (CoolLED PE-300 Ultra) through a dichroic mirror combined with an appropriate set of excitation and emission filters. Prior to each experiment, a small amount of fluorescent microspheres (FluoSpheres Invitrogen, diameter $1~$\textmu m, excitation/emission 505/515 nm) is dispersed in the water-glycerol mixture at a concentration of approximately 100 particles\,\textmu L$^{-1}$. The sample is then sonicated for 15 min in an ultrasonic bath to break up aggregates and ensure a homogeneous dispersion. These fluorescent particles serve as passive tracers for monitoring the internal flow trajectories within the drying droplet. Particle tracking velocimetry is performed using standard tracking algorithms (Python package \texttt{trackpy}) enabling the detection of individual tracer positions in each frame and the computation of the time-dependent radial velocity $v_r(r,z,t)$. Because the flow is three-dimensional, while the focal plane is fixed and the objective

\begin{figure}[ht!]
\includegraphics{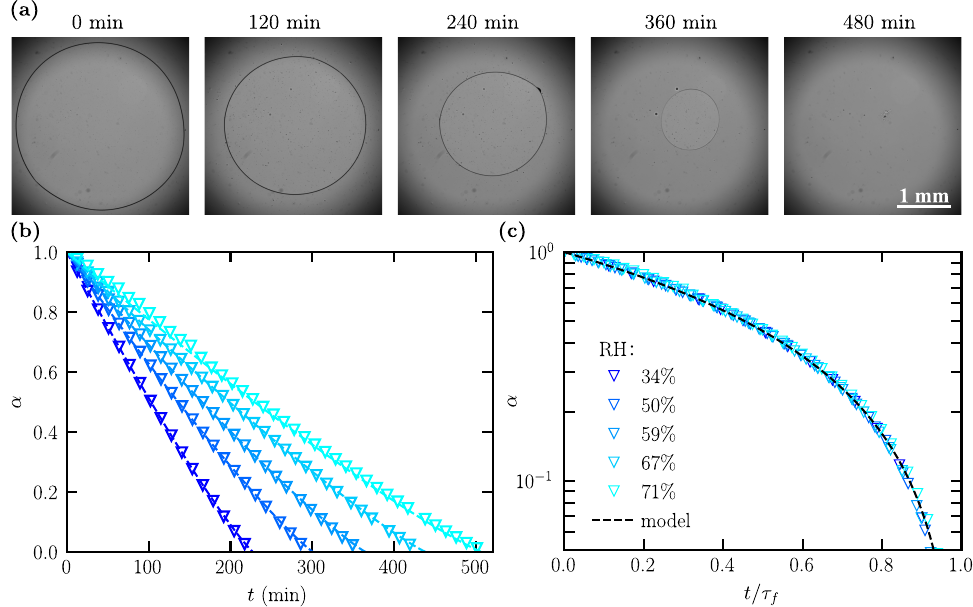}
\caption{\label{fig:drying_kinetics_water} Drying kinetics of a confined 2D pure water droplet  under different $\textrm{RH}$ levels. (a) Typical sequence of BF images for $R_0=1.56\textrm{ mm}$ and $\textrm{RH}=67\%$ (see also movie M1.avi in Ref.~\cite{Supp_Mat}). (b) Evolution of the normalized area $\alpha= A(t)/A_0$ over time for various $\textrm{RH}$ values. (c) Evolutions of $\alpha$ plotted against the normalized time $t/\tau_f$. Datasets collapse onto a single master curve in agreement with the analytical solution provided in Eq.~\eqref{equ:analytical}.}
\end{figure}

\noindent has a low numerical aperture, particles are tracked over almost the entire cell thickness but without precise information on their vertical position $z$. Therefore, when analyzing velocities, only the maximum values $v_{r,\rm{max}}(r,t)=\textrm{max}[v_r(r,z,t)]$ are considered for comparison with theory (see Sec.~\ref{sec:Internal_Flow}).  

When only BF imaging is used, images are recorded every 10~s. For experiments including fluorescence, both imaging modes are combined. The camera and illumination sources are then synchronized via a hardware trigger, enabling near-simultaneous acquisition of BF and fluorescence images every 5~s.

\section{\label{sec:RD}RESULTS AND DISCUSSION}

\subsection{Drying kinetics and observed phenomenology}

\subsubsection{Pure water droplet}

Figure~\ref{fig:drying_kinetics_water}(a) shows a typical BF observation of the drying of a pure water droplet in the confined 2D geometry, used here as a reference. Generally speaking, the droplet remains cylindrical for most of the drying process, until it disappears completely. In this experiment in which $R_0 =1.56$~mm and RH$=67\%$, the droplet dries in a total time $\tau_f=565$~min. As explained in

\begin{figure}[ht!]
\includegraphics{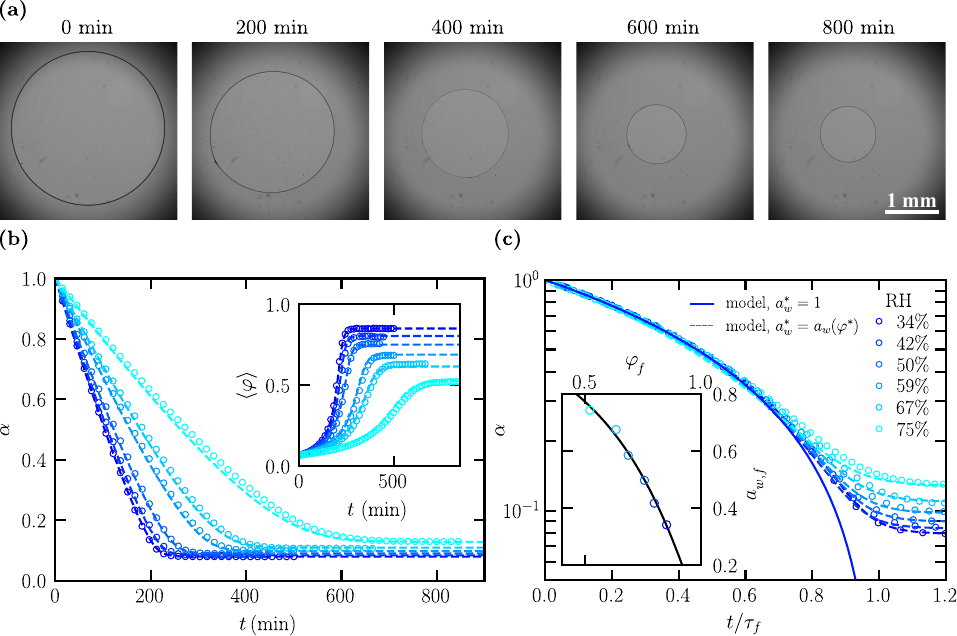}
\caption{\label{fig:drying_kinetics} 
Drying kinetics of a confined 2D water-glycerol droplet under different $\textrm{RH}$ levels. (a) Typical sequence of BF images for $R_0=1.54~\textrm{mm}$, $\varphi_0=0.0673$,  and $\textrm{RH}=75\%$ (see also movie M2.avi in Ref.~\cite{Supp_Mat}). (b) Evolution of the normalized area $\alpha$
over time for various $\textrm{RH}$ values. Inset: Corresponding evolutions of the mean glycerol volume fraction $\langle\varphi\rangle$ over time. (c) $\alpha$ vs normalized time $t/\tau_f$. Inset: Final glycerol volume fraction $\varphi_f$ vs the equilibrium water activity $a_{w,f}=a^{*}_{w}(t\to\infty)$, equal to the imposed external RH. The solid line corresponds to the water chemical activity of the mixture given by Eq.~\eqref{equ:bouchaudy_activity}. In panels (b) and (c), experimental data (blue symbols) are compared with the theoretical model from Sec.~\ref{sec:Model} (dashed lines).}
\end{figure}

\noindent Sec.~\ref{sec:MM}, a simple image analysis of the BF sequence allows extraction of the temporal evolution of the droplet area $A(t)$. The corresponding normalized area $\alpha = A(t)/A_0$ is plotted as a function of time in Fig.~\ref{fig:drying_kinetics_water}(b) for various RH levels, where $A_0=A(t=0)$ is the initial droplet surface area. As expected, increasing RH leads to a slower drying rate, i.e., a larger drying time $\tau_f$. By rescaling time with the experimentally measured $\tau_f$, Fig.~\ref{fig:drying_kinetics_water}(c) shows that all data collapse onto a single master curve, in excellent agreement with  Eq.~\eqref{equ:analytical}, the analytical solution of  the quasisteady, isothermal, diffusion-controlled drying model for pure water ($a_{w}^{*} = 1$). In addition, a comparison between the experimental values and theoretical predictions of $\tau_f$ is provided in  Supplemental Material~\cite{Supp_Mat}. The overall agreement confirms both the validity of the theoretical framework describing the drying dynamics of a pure water droplet in the 2D confined geometry and the robust performance of the custom-built humidity chamber.

\subsubsection{Water-glycerol droplet}

Similarly, Fig.~\ref{fig:drying_kinetics}(a) shows a typical BF observation of the drying of a water-glycerol mixture droplet in the confined 2D geometry, with an initial glycerol volume fraction $\varphi_0 = 6.73\%$, under comparable conditions.  While the droplet remains axisymmetric throughout the drying process, it does not evaporate completely but instead reaches a final state characterized by a radius $R_f=R(t\to \infty)\neq 0$. This behavior arises because the water chemical activity of the water-glycerol mixture differs from unity (that of pure water), resulting in a final equilibrium state with the surrounding atmosphere when $a_{w}^* = \mathrm{RH}$, as described by Eq.~\eqref{equ:evolution_alpha}.

The corresponding normalized area $\alpha$ is plotted as a function of time in Fig.~\ref{fig:drying_kinetics}(b), along with experiments performed at various RH levels. The drying kinetics initially follow a regime similar to that of a pure water droplet, as the chemical activity $a_{w}^*$ remains close to unity over a wide range of small glycerol volume fractions [see Fig.~\ref{fig:prop_gly_water}(c)]. From this first phase, the characteristic drying time $\tau_f$ can be extracted and is in good agreement with the theoretical predictions given by Eq.~\eqref{eq:tau_f} (see Supplemental Material~\cite{Supp_Mat}). Subsequently, the droplet gradually transitions toward its final equilibrium state, which depends on RH. As expected, higher RH values result in slower drying kinetics and larger final droplet areas [see Fig.~\ref{fig:prop_gly_water}(c)]. This behavior is further confirmed in the dimensionless representation shown in Fig.~\ref{fig:drying_kinetics}(c), in which time is normalized by the experimental $\tau_f$, and the model for pure water is superimposed. All datasets collapse onto the same curve during the initial phase, indicating identical evaporation kinetics, in good agreement with the quasisteady, isothermal, diffusion-controlled evaporation model for a pure water droplet ($a_{w}^*=1$). The curves then deviate in the second stage, corresponding to the  transition to the final equilibrium state when $a_{w}^* \neq 1$.
 
From $\alpha(t)$, the mean glycerol volume fractions $\langle\varphi(t)\rangle$ can be inferred using Eq.~\eqref{eq:avg_phi_ot}, as shown in the inset of Fig.~\ref{fig:drying_kinetics}(b). The results clearly reveal a monotonic increase in $\langle\varphi(t)\rangle$ from its initial value  $\varphi_0$ to the final equilibrium value $\varphi_f=\varphi(t\to\infty)$, which depends on the ambient RH. These findings, reported in the inset of Fig.~\ref{fig:drying_kinetics}(c), show excellent agreement with the expression of the activity $a_w(\varphi)$ given in Eq.~\eqref{equ:bouchaudy_activity}. This agreement further confirms the robustness  and stability of the custom-built humidity chamber.

\subsection{Concentration fields in the drying drop}

\begin{figure}[ht!]
\includegraphics{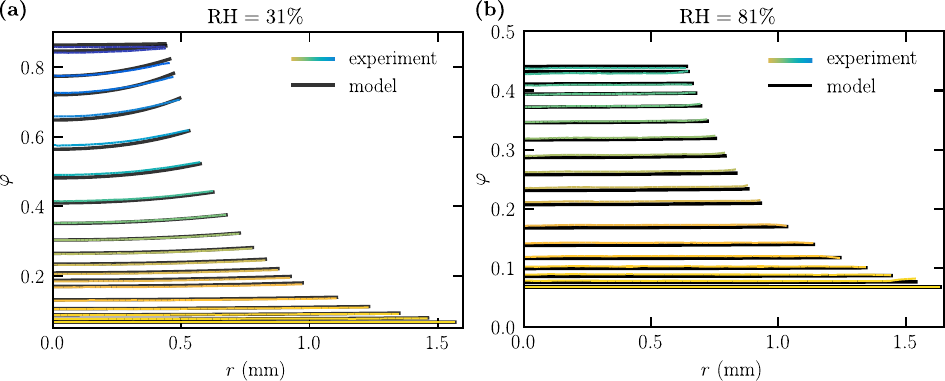}
\caption{\label{fig:profiles_comparison} Concentration profiles $\varphi(r,t)$ in a 2D confined water-glycerol droplet drying at two RH levels, with $\varphi_0 = 0.0673$. (a) $\textrm{RH}=31\%$, $R_0=1.57$ mm, time interval 10 min. (b) $\textrm{RH}=81\%$, $R_0=1.63$ mm, time interval 50 min. 
Experimental measurements obtained by interferometry (colored lines) are compared with the theoretical model described in Sec.~\ref{sec:Model} under the same conditions (black lines).}
\end{figure}

From the temporal sequence of interferograms, 2D glycerol volume fraction fields $\varphi(r,\theta,t)$ are extracted and, owing to axisymmetry, converted into angularly averaged profiles $\varphi(r,t)$ following the procedure described in Sec.~\ref{sec_MZI}.
The evolution of the glycerol volume fraction profiles for two experiments performed at markedly different relative humidity values, $\textrm{RH} = 31$ and $81\%$, is shown in Fig.~\ref{fig:profiles_comparison}. At low RH, corresponding to faster drying kinetics, slight concentration gradients develop, with higher glycerol volume fractions near the evaporation-induced receding meniscus. In contrast, at high RH, corresponding to slower drying kinetics, no significant gradients are observed, and the glycerol volume fraction increases almost uniformly within the drying droplet. More precisely, for $\textrm{RH} = 31\%$, the concentration profiles reveal that the droplet is initially homogeneous. At short times, a diffuse boundary layer forms near the receding meniscus and gradually extends toward the droplet center. At longer times, weak concentration gradients persist along the droplet radius until the final stage, when the composition becomes uniform again as the droplet reaches the final equilibrium state. This behavior is also visible in movies M3.avi and M4.avi provided in Supplemental Material~\cite{Supp_Mat}.

\subsection{Measurements of the diffusion coefficient and chemical activity}
\label{sec:Measurements of diffusion}

Assuming that mass transport in the drying droplet is governed primarily by diffusion [see Sec.~\ref{sec:Model}, particularly Eq.~\eqref{eqn:DeffDefinition}, and its validation in Sec.~\ref{sec:Internal_Flow}], the mutual diffusion coefficient of the mixture can be extracted from the interferometric measurements of $\varphi(r,t)$. To estimate precise values of $D(\varphi)$, the approach proposed in Ref.~\cite{sobac_collective_2020} is adopted, based on the global solute conservation equation obtained from spatial integration of Eq.~\eqref{eq:diff}:

\begin{equation}
    \frac{\partial \Psi(r, t)}{\partial t}= r D(\varphi) \frac{\partial \varphi}{\partial r},
\label{eq:diff_int}
\end{equation}
which relates the diffusive flux at a given radius $r< R(t)$, to the temporal variation of
\begin{equation}
    \Psi(r, t) = \int_{0}^{r} u\, \varphi(u, t) \,\text{d}u.
\label{eq:psi}
\end{equation}
All quantities in Eq.~\eqref{eq:diff_int} can be computed directly from experimental measurements of $\varphi(r,t)$ at a given  $r$, yielding $D$ values corresponding to the local volume fraction $\varphi(r,t)$.

\begin{figure}[ht!]
\includegraphics{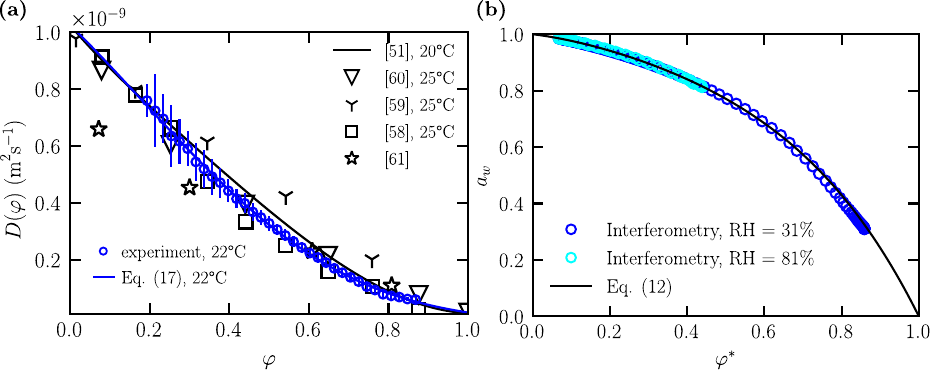}
\caption{\label{fig:diffusion_activity}Measured mutual diffusion coefficient $D(\varphi)$ and water chemical activity $a_w(\varphi)$ for the water-glycerol mixture at $T=22^{\circ}$C. (a) $D(\varphi)$ estimated from the measurements of $\varphi(r,t)$ using Eq.~\eqref{eq:diff_int}. The error bars are calculated using the standard deviation over the equivalent of 30 experimental curves of $D(\varphi)$. The blue line is a fit by Eq.~\eqref{eq:fitD}, with literature data (black symbols) included for comparison~\cite{derrico_diffusion_2004, ternstrom_mutual_1996, nishijima_diffusion_1960, rashidnia_measurement_2004, bouchaudy_steady_2018}. (b) $a_w(\varphi)$ at the interface $\varphi^*$, estimated from the measurements of $\varphi(r,t)$ at two relative humidities, $\textrm{RH}=31$ and 81\%, using  Eq.~\eqref{equ:evolution_alpha}. The solid line corresponds to Eq.~\eqref{equ:bouchaudy_activity}.}
\end{figure}

Figure~\ref{fig:diffusion_activity}(a) shows the mutual diffusion coefficient of the water-glycerol mixture computed from ten radial positions across three distinct experiments with similar initial conditions at low $\textrm{RH}=31\%$. The mean value of $D(\varphi)$ obtained from all measurements is plotted, with error bars indicating the standard deviation calculated over the 30 points of measurement. The verified reproducibility of the results, combined with the small error bars, strongly supports the conclusion that solute transport within the drying drop is accurately described by the diffusion equation Eq.~\eqref{eq:diff}, with the mutual diffusion coefficient $D(\varphi)$ shown in Fig.~\ref{fig:diffusion_activity}(a). Remarkably, $D(\varphi)$ can thus be obtained from a single experiment at low RH, covering nearly the entire volume fraction range.

The diffusion coefficients measured here are in good agreement with literature values~\cite{derrico_diffusion_2004,ternstrom_mutual_1996,nishijima_diffusion_1960,rashidnia_measurement_2004,bouchaudy_steady_2018}, although slight differences may arise due to variations in experimental temperatures ($T=20$--$25^\circ \textrm{C}$) or unspecified conditions.
Based on our experimental data, a fourth-degree polynomial is proposed for $D(\varphi)$ for the water-glycerol mixture at $T = 22^\circ$C:
\begin{equation}
D(\varphi) = (10.25 - 13.505\, \varphi - 10.037\, \varphi^2 + 22.431\, \varphi^3 - 8.998\, \varphi^4) \times 10^{-10}~\textrm{m$^2$\,s$^{-1}$}\, ,
\label{eq:fitD}
\end{equation}
yielding an excellent reproduction of the experimental data, even when $\varphi \to 1$. Note that the value at infinite dilution $D(\varphi \to 0) = 10.25 \times 10^{-10}$~m$^2$\,s$^{-1}$ from Ref.~\cite{derrico_diffusion_2004} was imposed for the fit.

We can also estimate the water chemical activity of the mixture against the concentration using our combined measurements of $\varphi^* = \varphi[R(t),t]$ and $\alpha(t)$ from interferometry. Indeed, $(a_w^* -\rm{RH})$ is the driving force for evaporation and governs the drying kinetics according to Eq.~\eqref{equ:evolution_alpha}. The experimental derivative of $\alpha(t)$ thus leads to the activity of the mixture, plotted against the measured values of the concentration at the interface $\varphi^*$, as shown in Fig.~\ref{fig:diffusion_activity}(b) for our two markedly different imposed RH values, in really good agreement with the water-glycerol expression of $a_w(\varphi)$ in Eq.~\eqref{equ:bouchaudy_activity}.

It is worth noting the advantages of performing experiments at low RH for extracting both $a_{w}(\varphi)$ and $D(\varphi)$. First, at low RH, drying proceeds to higher final volume fractions due to the shape of $a_{w}(\varphi)$, providing measurements over a larger range of $\varphi$: $0.2$--$0.9$ for $D(\varphi)$ and $0.05$--$0.9$ for $a_{w}(\varphi)$ (see Fig.~\ref{fig:diffusion_activity}). Second, at low RH, concentration gradients are sufficiently large, yielding physically meaningful and more accurate values of $D(\varphi)$. When gradients are too weak or absent, the present methodology based on Eq.~\eqref{eq:diff_int} results in nonphysical values. These observations are fully consistent with the estimated P\'eclet numbers [Eq.~\eqref{eq:Pe}], which are $\mathrm{Pe}\simeq 0.18$ at RH = 31\% and $\mathrm{Pe}\simeq 0.05$ at RH = 81\%, corresponding respectively to steeper and smoother concentration profiles.
Note that concentration gradients depend only weakly on the geometrical parameters $R_0$ and  $R_{\textrm{W}}$ [see the logarithmic term in the P\'eclet number definition  Eq.~\eqref{eq:Pedevelop}]. More pronounced gradients could develop for $R_0 \lesssim  R_\textrm{W}$, but this configuration lies outside the scope of the approximations used in the present study ($R_0 \ll R_\textrm{W}$).

Thanks to our combined global and local analysis of the drying kinetics [$\alpha(t)$ and $\varphi(r,t)$], we managed to estimate the water chemical activity of the mixture and its mutual diffusion coefficient against concentration. We now illustrate the self-consistency of our approach by solving the model describing the drying dynamics [Eqs.~\eqref{equ:evolution_alpha}, \eqref{eq:diff}, and \eqref{eq:nonvo_bc}] using our experimental estimates of $a_w(\varphi)$ and $D(\varphi)$. This allows us to compute the evolutions of the concentration field $\varphi(r,t)$ and of $\alpha(t)$, and to compare them to our measurements. Figures~\ref{fig:drying_kinetics} and \ref{fig:profiles_comparison} show the comparison between the full experimental data and the numerical results of the model, revealing good agreement up to complete drying, for the full range of RH investigated, without any adjustable parameters. This validates our experimental approach, the precision of the mixture characterization, and our comprehensive understanding of the drying process and the associated mechanisms at play.

\subsection{Internal buoyancy-driven convection: Present but negligible}
\label{sec:Internal_Flow}

\begin{figure}[ht!]
\includegraphics{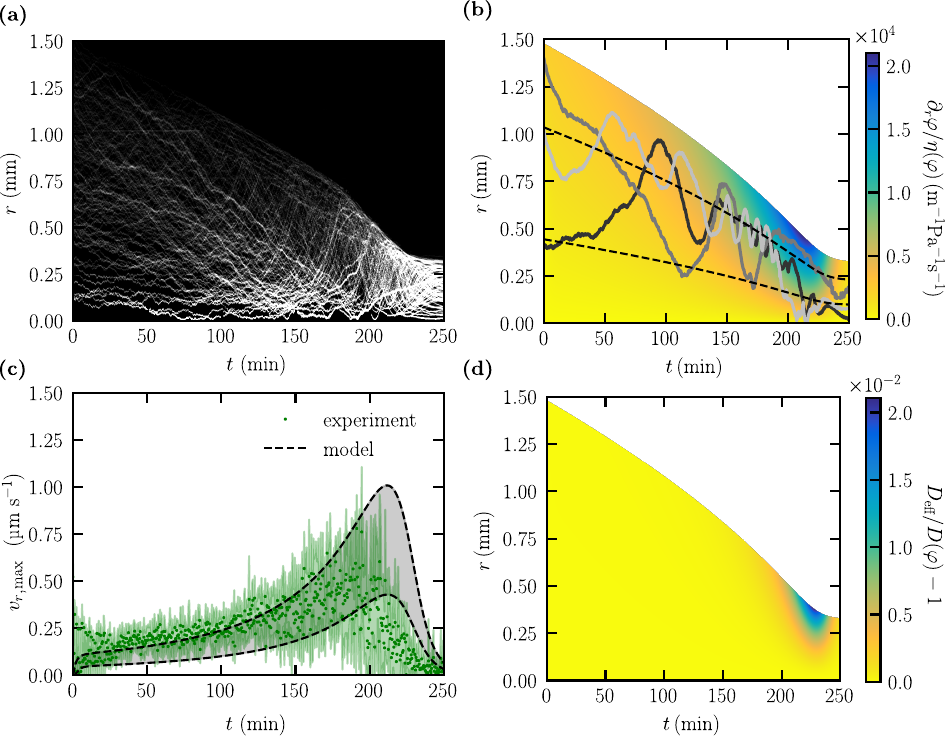}
\caption{\label{fig:convection_condition2} Internal flow motion and mass transport in a confined 2D drying droplet of a water–glycerol mixture with $R_0 = 1.48$~mm,  $\varphi_0 = 0.0392$, and $\text{RH} = 40\%$. (a) Spatiotemporal diagram evidencing the traces of dispersed particles via fluorescence microscopy (see also movie M5.avi in Ref.~\cite{Supp_Mat}). (b) Corresponding theoretical spatiotemporal diagram of $\partial_r \varphi / \eta(\varphi)$. Experimental trajectories of three fluorescent microparticles from panel (a) are superimposed as solid lines. (c) Comparison between the maximum radial velocities $v_{r,\mathrm{max}}$ obtained from particle tracking between $r = 0.3R(t)$ and $0.7R(t)$ [see dashed lines in (b)], and the predictions of  Eq.~\eqref{equ:max_rad_velocity} (gray area). (d) Corresponding theoretical spatiotemporal diagram of the dimensionless criterion $[D_{\mathrm{eff}}/D(\varphi)]-1$ defined in Eq.~\eqref{eqn:DeffDefinition}, computed for the same experimental conditions.}
\end{figure}

The above analysis suggests that solute transport within the confined water-glycerol droplet is correctly described by the transport equation Eq.~\eqref{eq:diff}, and therefore dominated by diffusion. However, as revealed by the experiment shown in Fig.~\ref{fig:convection_condition2}(a) ($\mathrm{RH}=40\%$, $R_0=1.48$~mm, $\varphi_0 =0.0392$),
convective motions are readily observed in the drying droplet when fluorescent particles are dispersed in the water-glycerol mixture (see also movie M4.avi in Ref.~\cite{Supp_Mat}). These radial recirculation flows are directed toward the center of the droplet near the bottom wafer and toward its edge near the top wafer [see arrows in Fig.~\ref{fig:model}(b)]. Such flows have already been observed during the drying of confined droplets of solutions or dispersions, and have been attributed to the radial density gradient induced by drying~\cite{daubersies_confined_2012,loussert_drying_2016,sobac_collective_2020,selva_solutal_2012}. This gradient is indeed orthogonal to the gravitational field $g$ and therefore associated with a buoyancy-driven flow. 
Computing theoretically these flows is  challenging  due to  variations in viscosity and possible coupling with mass transport. 
Nevertheless, within the framework of the lubrication approximation and assuming a uniform viscosity $\eta$, the velocity profile resulting from a radial density gradient is~\cite{selva_solutal_2012}
\begin{equation}
v_{r}(r,z,t) \simeq  \frac{\rho_{\textrm{w}}\beta  g}{12 \eta}  z(z-h)(2z-h)\frac{\partial\varphi}{\partial r},
\label{equ:lubrification}
\end{equation}
with maximal values at the planes $z \simeq 0.2 h$ and $\simeq 0.8 h$ given by
\begin{equation}
v_{r, \textrm{max}}(r,t) \simeq 0.008 \frac{\rho_{\textrm{w}}\beta g h^3}{\eta}  \frac{\partial\varphi}{\partial r}.
\label{equ:max_rad_velocity}
\end{equation}
A direct comparison between fluorescence microscopy experiments and predictions of Eq.~\eqref{equ:lubrification}, as done for instance in Ref.~\cite{selva_solutal_2012}, is not trivial here because the low numerical aperture objective   precludes us from determining the exact vertical position 
$z$ of the tracked particles within the droplet. Nevertheless, we extracted trajectories of $\simeq$50~particles using a standard particle tracking algorithm during the drying of the droplet [see Fig.~\ref{fig:convection_condition2}(b) for only a few]. These trajectories are superimposed to the map of $\partial_r \varphi / \eta(\varphi)$ as suggested by Eq.~\eqref{equ:max_rad_velocity}, with the concentration field $\varphi(r,t)$ predicted by model given by Eqs.~\eqref{equ:evolution_alpha}, \eqref{eq:diff}, and \eqref{eq:nonvo_bc} for the conditions explored in this experiment, and $\eta$ estimated from the local  concentration $\varphi(r,t)$ using the relation shown in Fig.~\ref{fig:prop_gly_water}(b). Radial velocities $v_r(r,z,t)$ are estimated by differentiating experimental trajectories on a temporal step of $15$~s  to minimize error location due to Brownian motion. Figure~\ref{fig:convection_condition2}(c) displays the average of the 12 maximal values of these velocities at each time $t$ with the condition $0.3R(t)<r<0.7R(t)$. Error bars are given by the standard deviation of these 12 values and are mainly due to Brownian motion. These estimated radial velocities should correspond to those of particles that are located close to the planes of maximal velocities, $z \simeq 0.2 h$ and
$\simeq 0.8 h$, i.e., $v_{r,\mathrm{max}}(r,t)$. These velocities vary with the drying of the droplet in a nonmonotonic manner, with values ranging from $v_{r,\mathrm{max}} \simeq 0.1$ to $\simeq 0.8~$\textmu  m\,s$^{-1}$.  These data, obtained for  $0.3R(t)<r<0.7R(t)$, in a spatial region for which the lubrication approximation holds (i.e., far from the center and edge of the droplet), are superimposed on the prediction of Eq.~\eqref{equ:max_rad_velocity}, showing good overall agreement. This confirms the origin of these flows and validates the above theoretical approach.
These flows  can \textit{a priori} impact mass transport within the droplet and lead to deviation from the diffusion equation Eq.~\eqref{eq:diff}.
Still within the framework of the lubrication approximation and again assuming a uniform viscosity $\eta$ and diffusivity $D$, Salmon and Doumenc ~\cite{salmon_buoyancy-driven_2020} showed using a Taylor-Aris-like approach that 
the solute concentration field $\varphi(r,t)$ still obeys a diffusive equation as Eq.~\eqref{eq:diff}, but with an effective dispersion coefficient $D_{\mathrm{eff}}$ given by
\begin{equation}
D_{\mathrm{eff}}  = D\left[1+ \frac{1}{362880} \left(\frac{\rho_{\textrm{w}}\beta g  h^4}{\eta D} \frac{\partial \varphi}{\partial r} \right)^2\right].
\label{eqn:DeffDefinition}
\end{equation}
The last term accounts for dispersion of the solute by the buoyancy-driven flow, and scales as $\sim$$(v_{r, \textrm{max}} h/D)^2$ as in the standard Taylor-Aris approach. Equation~\eqref{eqn:DeffDefinition} can be used to evaluate the role of free convection on  solute transport, and diffusion \textit{a priori} dominates as soon as the last term is small compared to $1$. Figure~\ref{fig:convection_condition2}(d) shows the last term of Eq.~\eqref{eqn:DeffDefinition} estimated theoretically from the model and empirical laws for $\eta(\varphi)$ and $D(\varphi)$. This 2D map shows that this term remains negligible throughout the drying of the droplet, with a maximal value of $\simeq$0.02 at $t \simeq 230$~min corresponding to $\langle \varphi \rangle \simeq 0.58$. This shows that the unavoidable buoyancy-driven flows due to drying, have no impact on mass transport within the drop, so that the diffusion equation Eq.~\eqref{eq:diff} can be used safely to describe the dynamics. We checked numerically that the same conclusion can be drawn for all experimental conditions explored in the present work (mainly the different $R_0$ and RH; see Figs.~\ref{fig:drying_kinetics} and \ref{fig:profiles_comparison}). These additional results are provided in the Supplemental Material~\cite{Supp_Mat}. Note, however, that because of the scaling $\sim$$h^8$ of the last term of Eq.~\eqref{eqn:DeffDefinition}, the same conclusion cannot be drawn for larger cell height, even $h=300$~\textmu m. Because of the nonlinear nature of the problem investigated [$a_w(\varphi)$, $D(\varphi)$, and $\eta(\varphi)$], numerical estimates using Eq.~\eqref{eqn:DeffDefinition} are necessary to assess the fact that free convection can be neglected for a given height $h$.

\section{CONCLUSIONS}

We have presented a methodology for finely characterizing the drying of complex fluids under precisely controlled conditions, combining Mach-Zehnder interferometry, a custom RH-controlled chamber, and a confined 2D droplet geometry. This integrated approach enables high-accuracy ($\pm 0.5\%$), spatially and temporally resolved ($6\,$\textmu m\,pixel$^{-1}$ and 1 frame\,s$^{-1}$, respectively) measurements of both drying kinetics and internal concentration fields over a wide RH range (25--95\%, with an accuracy $\pm 3.5\%$).

Applied to the water-glycerol binary mixture, the method reveals how the drying rate, the final equilibrium state, and the internal concentration fields depend on RH, through the driving factor [$a_w(\varphi)$-RH]. Increasing RH (and thus decreasing the P\'eclet number) leads to slower evaporation, smoother concentration gradients, and larger final droplets at lower equilibrium concentrations. All experimental measurements are well predicted by a quasisteady, isothermal model of vapor-diffusion-controlled evaporation coupled to solute diffusion inside the droplet \cite{daubersies_evaporation_2011,sobac_collective_2020}, demonstrating that drying in this confined geometry is thoroughly controlled and quantitatively understood.

Remarkably, the combined use of drying kinetics $\text{d}R/\text{d}t$ and the spatiotemporal concentration field $\varphi(r,t)$ enables the simultaneous extraction of both the mutual diffusion coefficient $D(\varphi)$ and the water chemical activity $a_w(\varphi)$ over nearly the full glycerol volume-fraction range, even from a single experiment at low initial $\varphi_0$ and RH. While our measurements of $a_w(\varphi)$ agree with literature data \cite{bouchaudy_steady_2018} and Eq.~(\ref{equ:bouchaudy_activity}), we provide a consistent fit for $D(\varphi)$ [see Eq.~(\ref{eq:fitD})].

Complementary fluorescence-microscopy measurements quantified the internal flow, showing its dependence on concentration gradients and confirming its consistency with unavoidable buoyancy-driven convection predicted by the lubrication-approximation model~\cite{salmon_buoyancy-driven_2020}. Although detectable, the flow remains weak (maximum velocities $\sim 1~$\textmu$\mathrm{m\,s^{-1}}$) and negligible to diffusion. The diffusion-dominated regime is validated throughout all investigated parameters by the condition $\{[D_{\mathrm{eff}}/D(\varphi)]-1\} < 0.1$.

The strong agreement between experiments, modeling, and literature data demonstrates that the present approach provides a powerful and quantitative framework for investigating evaporation and mass transport in complex fluids. The confined droplet geometry allows precise manipulation of small volumes, easy optical observation, and well-defined drying conditions free from contact-line and interfacial complexities, while suppressing Rayleigh-Bénard-Marangoni flows, thus enabling a simple one-dimensional diffusive description of the problem. This confined drying approach is further enhanced by RH control and interferometric characterization, providing high accuracy and spatial-temporal resolution. Although demonstrated here on the water-glycerol system, the methodology is broadly applicable to a wide range of complex fluids, including highly viscous or polymer solutions, colloidal dispersions, etc. By exploiting evaporation as an out-of-equilibrium route from dilute to dense regimes, this method allows precise determination of water chemical activity and mutual diffusion coefficients—parameters that are often difficult to measure in concentrated systems using conventional techniques. Together, these features establish a versatile and generalizable framework for investigating drying dynamics and mass transport in complex fluids under rigorously controlled thermodynamic and geometric conditions.

\section*{ACKNOWLEDGEMENTS}
The authors gratefully acknowledge B. Niez and M. Luthy, for their technical support in the development and implementation of the humidity-controlled chambers, and G. Clisson for the preparation of the coated wafers. We also thank T. Burgaud, B. Fagnot, and P. Hayet for their assistance with preliminary experiments carried out during their undergraduate internships. This research was financially supported by the R\'egion Nouvelle-Aquitaine and the Universit\'e de Pau et des Pays de l’Adour through the AMIBES project. B.S. also acknowledges support from the French National Centre for Scientific Research (CNRS), the Energy and Environment Solutions program of the Université de Pau et des Pays de l’Adour (E2S-UPPA), and the French National Centre for Space Studies (CNES).

\section*{DATA AVAILABILITY}
The data that support the findings of this article are not publicly available. The data are available from the authors upon reasonable request.

\newpage 

\section*{APPENDIX}
\appendix
\section{NUMERICAL RESOLUTION OF THE THEORETICAL MODEL}
\label{sec:App_Model}

To numerically solve the model described in Sec.~\ref{sec:Model}, the following dimensionless variables are introduced:
\begin{eqnarray}
   &&\tilde{r}=r/R_0, \\
   &&\tilde{R}=R/R_0, \\
   && \tilde{t} = t/\tau_d,\\ 
   &&\hat{D}(\varphi)= D(\varphi)/D_0,
\end{eqnarray}
with $\tau_d$ given by  
\begin{equation}
    \tau_d = \frac{R_{\textrm{0}}^2}{4 D_{w} V_{w} C_{\textrm{sat}} } [1 - \textrm{ln}(\beta)]\, ,
\end{equation}
the characteristic drying time of a pure water droplet at $\textrm{RH} = 0\%$.
With these dimensionless variables, Eqs.~\eqref{equ:evolution_alpha}, \eqref{eq:diff} and \eqref{eq:nonvo_bc}
now read
\begin{eqnarray}
&&\frac{\textrm{d}\alpha}{\textrm{d}\tilde{t}} = \frac{1 - \ln(\beta)}{\ln(\beta \alpha)} (a_w\{\varphi[\tilde{R}(\tilde{t}),\tilde{t}]\} - \textrm{RH}), \label{eq:dadtunitless}\\
    &&\frac{\partial \varphi}{\partial \tilde{t}} = \frac{1}{\tilde{r}} \frac{\partial}{\partial \tilde{r}} \left( \tilde{r} \frac{\hat{D}(\varphi)}{\mathrm{Pe}} \frac{\partial \varphi}{\partial \tilde{r}} \right)
\label{eq:diff_dl},\\
&&-\hat{D}(\varphi)\frac{\partial \varphi}{\partial \tilde{r}}\bigg|_{\tilde{r}=\tilde{R}(\tilde{t}),\tilde{t}} = \varphi[\tilde{R}(\tilde{t}),\tilde{t}]\frac{\textrm{d}\tilde{R}}{\textrm{d}\tilde{t}}\, .
\label{eq:bd_dl}
\end{eqnarray}

With the change of variable $\varphi(\tilde{r},\tilde{t}) \rightarrow \varphi(u,\tilde{t})$ with $u = \tilde{r}/\sqrt{\alpha}$,
Eqs.~\eqref{eq:diff_dl} and \eqref{eq:bd_dl} become
\begin{eqnarray}
    &&\frac{\partial(\alpha\varphi)}{\partial\tilde{t}} = \frac{1}{u}\frac{\partial}{\partial u} \left(  u \left[ \frac{\hat{D}(\varphi)}{\mathrm{Pe}} \frac{\partial \varphi}{\partial u} + \frac{u}{2} \varphi \frac{\textrm{d}\alpha}{\textrm{d}\tilde{t}} \right] \right),\label{eq:unitless1}\\
    &&\left[ \frac{\hat{D}(\varphi)}{\mathrm{Pe}} \frac{\partial \varphi}{\partial u} + \frac{u}{2} \varphi \frac{\textrm{d}\alpha}{\textrm{d}\tilde{t}} \right]_{u=1, \tilde{t}} = 0.\label{eq:unitless2}
\end{eqnarray}
Equations \eqref{eq:dadtunitless}, \eqref{eq:unitless1}, and \eqref{eq:unitless2} are finally solved using Python over the fixed space interval $u \in [0$--$1]$, with the initial condition $\varphi(u, \tilde{t}=0) = \varphi_0$.

\section{INFLUENCE OF PDMS-COATING THICKNESS ON CONFINED DRYING EXPERIMENTS\label{ss:influence_of_PDMS}}

\begin{figure}[ht!]
\includegraphics{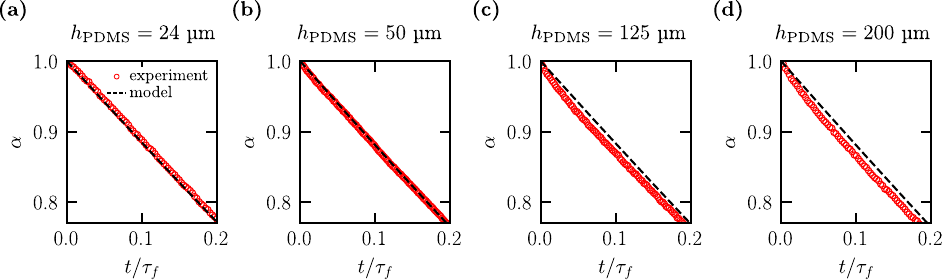}
\caption{\label{fig:PDMS_pumping} Influence of the thickness of the  PDMS coatings $h_\text{PDMS}$ on the drying kinetics of a water droplet. (a–d) Normalized droplet area $\alpha$ against normalized time
$t/\tau_f$ at small times,  for increasing $h_\text{PDMS}$.}
\end{figure}

To assess the  possible influence of the PDMS coatings, we investigated the drying of pure water droplets under nearly identical conditions ($R_0 = 1.2 - 1.5$~mm, $\text{RH} = 0.35 - 0.6$), but using glass wafers coated with PDMS layers of varying thicknesses. Figure~\ref{fig:PDMS_pumping} reports the measured normalized droplet area $\alpha$ vs\ normalized time $t/\tau_f$, along with the fits given by Eq.~\eqref{equ:analytical}. These data show that the \, theoretical prediction perfectly accounts for the kinetics \, for the  thinnest

\noindent coatings ($h_\text{PDMS} = 24$ and  $50~$\textmu m), consistent with the results reported in Fig.~\ref{fig:drying_kinetics_water}. However, for the thickest coatings ($h_\text{PDMS} = 125$ and  $200~$\textmu m), the kinetics slightly deviate from Eq.~\eqref{equ:analytical}. More specifically, $\alpha$ decreases more rapidly than predicted on timescales $t\simeq 0.01\tau_f$ ($1-3$~min), before following the kinetics predicted by Eq.~\eqref{equ:analytical} but shifted from a value of the order of $\simeq$0.03. This suggests that a significant fraction of the initial droplet volume ($\simeq$0.03) may be adsorbed into the PDMS layers in these cases.

Harley \textit{et al.}~\cite{Harley2012} reported a comprehensive study via gravimetric analysis on  water vapor sorption in PDMS, specifically  Sylgard 184 as in the present work. The moisture content  $C_\text{PDMS}$ (mol\,m$^{-3}$) depends at equilibrium on the water chemical activity $a_w$ of the vapor (equivalently the residual humidity $\text{RH}$), in contact with PDMS, and follows various modes of sorption. This study shows, as a rough first approximation, that  $C_\text{PDMS}  \simeq a_w C_\text{PDMS}^\text{sat}$, with $C_\text{PDMS}^\text{sat} \simeq 40~$mol\,m$^{-3}$. The same authors also studied the dynamics of vapor sorption and reported the diffusion coefficient of water in PDMS in the range $D_\text{PDMS} = 5-10 \times 10^{-10}$~m$^2$\,s$^{-1}$ depending on the water content. 
In the 2D configuration of the drying droplet shown in Fig.~\ref{fig:model}, there is a radial flux of water  in the gas phase [see Eq.~\eqref{eq:aw_r}], and thus also a radial concentration profile of water within the PDMS coatings. Because  $D_w  C_\textrm{sat} \gg C_\text{PDMS}^\text{sat} D_\text{PDMS}$, transport of water through the gas phase dominates that through PDMS, and  water evaporation is described by the model given in Sec.~\ref{sec:Model}. However, the amount of water within the PDMS matrix may not be negligible and could explain the  deviations shown in Fig.~\ref{fig:PDMS_pumping}. 
Assuming local equilibrium between the vapor phase and the PDMS coatings, the moisture content within PDMS in the case of a droplet of pure water is given by 
\begin{eqnarray}
    && C_\text{PDMS}(r,t)=C_\text{PDMS}^\text{sat}~\text{for}~~r\leq R(t)~~\text{and} \notag \\ 
     &&C_\text{PDMS}(r,t) = (C(r,t)/C_\mathrm{sat})C_\text{PDMS}^\text{sat}~\text{for}~~R(t)<r\leq R_\mathrm{W},
    \label{eq:concwaterpdms}
\end{eqnarray}
 with $C(r,t)$ given by Eq.~\eqref{eq:aw_r},
 the term $C(r,t)/C_\mathrm{sat}$ being the profile of  water chemical activity of the vapor in the cell.
For PDMS coatings that are initially in equilibrium with the vapor at ambient $\text{RH}$ (i.e., $C_\text{PDMS} = \mathrm{RH}\, C_\text{PDMS}^\text{sat}$), 
integration of Eq.~\eqref{eq:concwaterpdms} over the whole cell can be used to estimate the volume of liquid water needed to reach the local equilibrium between the vapor phase and the PDMS matrix. 
For $R = R_0 \ll R_w$, the ratio of this volume to the droplet volume $\pi R_0^2 h$  is given after calculation  by
\begin{equation}
    \mathcal{R} \simeq - 2 
    C_\text{PDMS}^\text{sat} V_m \frac{h_\mathrm{PDMS}}{h}\frac{1-\text{RH}}{\beta\ln(\beta)}. \label{eq:RatioPDMS}
\end{equation}
In the conditions explored in Fig.~\ref{fig:PDMS_pumping}, $\mathcal{R} \ll 1$ for $h_\mathrm{PDMS} = 24~$\textmu m, but reaches a value of the order of $\mathcal{R} \simeq 0.1$ for $h_\mathrm{PDMS} = 200~$\textmu m, demonstrating that a significant part of the initial volume of the droplet is absorbed within the PDMS coatings at early times to reach local equilibrium. The values of the shift observed in Fig.~\ref{fig:PDMS_pumping} for the thickest coatings ($\simeq$0.03) are slightly smaller than the  predictions given by Eq.~\eqref{eq:RatioPDMS}. This is possibly due to the amount of water absorbed by the PDMS coatings just before the initial start of the experiment ($t=0$). The duration of the experiment preparation period, of the order of $\simeq 30$~s, is indeed comparable to the diffusion timescale of water in PDMS, even for the thickest coating, $h_\mathrm{PDMS}^2/D_\text{PDMS} \simeq 1$~min.  Further experiments would be needed to deepen the comparison between experiments and models.
Despite the relatively coarse assumptions of our model, Eq.~\eqref{eq:RatioPDMS} allows one to estimate the maximum thickness of the PDMS coatings to avoid these subtle effects.

While PDMS coatings promote hydrophobic conditions and enable axisymmetric droplet shrinkage without pinning in confined drying experiments, water absorption by PDMS can significantly affect drying kinetics depending on its thickness. To keep its influence negligible compared to the evaporation dynamics modeled in Sec.~\ref{sec:Model}, the coating should be sufficiently thin.

\bibliography{2DdryingGlycerolwaterpaper}

\end{document}